\documentclass{PoS}
\usepackage{sidecap}

\title{Trinity: An Air-Shower Imaging System for the Detection of Ultrahigh Energy Neutrinos}

\ShortTitle{Trinity -- A UHE neutrino air-shower imaging system}

\author{\speaker{Nepomuk Otte}$^a$, Anthony M. Brown$^b$, Abraham D.\ Falcone$^c$, Mos\`e Mariotti$^d$, and Ignacio Taboada$^a$\\
\llap{$^a$}School of Physics \& Center for Relativistic Astrophysics, Georgia Institute of Technology,\\\ 837 State Street NW, Atlanta, GA 30332-0430, USA\\
\llap{$^b$} Center for Advanced Instrumentation, Department of Physics, Durham University, South Road, Durham, UK, DH1 3LE\\
\llap{$^c$} Department of Astronomy \& Astrophysics, Pennsylvania State University,
University Park,\\ PA 16802, USA  \\
\llap{$^d$} Dipartimento di Fisica e Astronomia, Universit\`a di Padova \& INFN Sezione di Padova\\ Via F. Marzolo 8, Padova, I-35132, ITALY  \\
E-mail: \email{otte@gatech.edu}}

\abstract{
Efforts to detect ultrahigh energy neutrinos are driven by several objectives: What is the origin of astrophysical neutrinos detected with IceCube? What are the sources of ultrahigh energy cosmic rays? Do the ANITA detected events point to new physics? Shedding light on these questions requires instruments that can detect neutrinos above $10^7$\,GeV with sufficient sensitivity -- a daunting task. While most ultrahigh energy neutrino experiments are based on the detection of a radio signature from shower particles following a neutrino interaction, we believe that the detection of Cherenkov and fluorescence light from shower particles is an attractive alternative. Imaging air showers with Cherenkov and fluorescence light is a technique that is successfully used in several ultrahigh energy cosmic ray and very-high energy gamma-ray experiments. 
We performed a case study of an air-shower imaging system for the detection of earth-skimming tau neutrinos. The detector configuration we consider consists of an imaging system that is located on top of a mountain and is pointed at the horizon. From the results of this study we conclude that a sensitivity of $3\cdot10^{-9}$\,GeV\,cm$^{-2}$s$^{-1}$sr$^{-1}$ can be achieved at $2\cdot10^8$\,GeV with a relatively small and modular system after three years of observation.
In this presentation we discuss key findings of our study and how they translate into design requirements for an imaging system we dub Trinity.}

\FullConference{36th International Cosmic Ray Conference -ICRC2019-\\
		July 24th - August 1st, 2019\\
		Madison, WI, U.S.A.}

\begin{document}

\section{Introduction}

The ultrahigh energy (UHE; $>10^7$\,GeV) neutrino band is an untapped resource when it comes to address some of the most interesting questions in astroparticle physics. At lower energies, IceCube has detected neutrinos of astrophysical origin \cite{Aartsen2013}. These astrophysical neutrinos are generated when cosmic rays interact in or around their acceleration site, but the identification of these sources is still a puzzle that remains largely unsolved. The recent correlated detection of neutrinos and very-high energy (VHE) gamma rays from TXS\,0506+056 points to blazars as one possible neutrino source \cite{IceCubeCollaboration2018a,IceCubeCollaboration2018b}, but more correlated detections are needed in order to draw firm conclusions. Observations in the UHE band will help to narrow in on source candidates, while providing more potential events for correlated detections. 

Ultrahigh energy cosmic rays (UHECRs) can also produce UHE neutrinos by interacting with the cosmic microwave background (CMB). The flux of these so-called cosmogenic neutrinos has imprinted information about the composition of UHECRs and their sources, \emph{e.g.}\,\cite{2019arXiv190306714A}. Furthermore, UHE neutrinos probe neutrino physics at the highest energies, which could potentially hint at new physics beyond the Standard Model, see \emph{e.g.}\,\cite{2019arXiv190700991B}, or superheavy dark matter, see \emph{e.g.}~\cite{2012JCAP...10..043M}. 

\begin{SCfigure}[1.0][t]
\includegraphics[width=.6\textwidth]{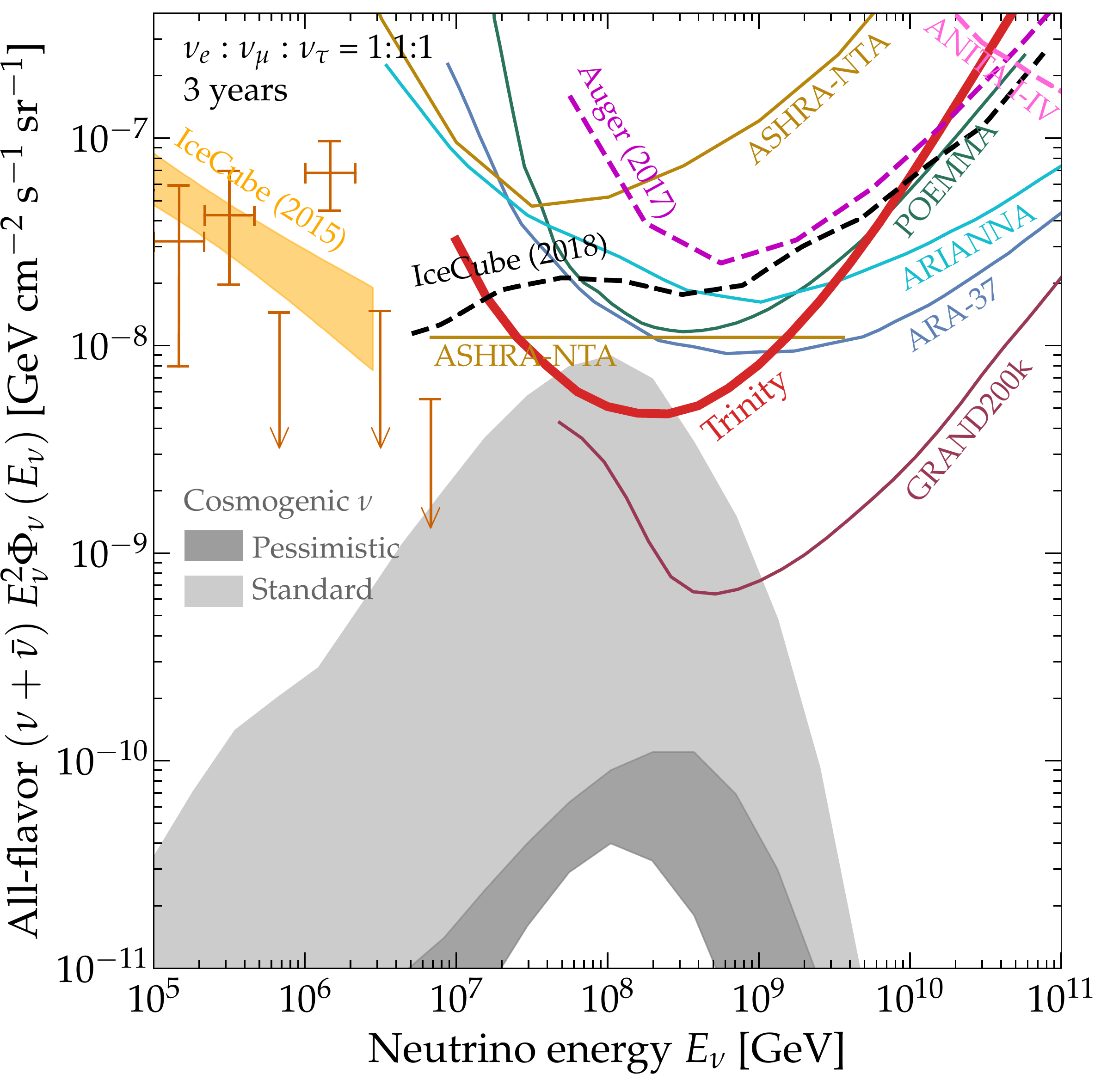}
\caption{All-flavor sensitivity of \emph{Trinity} (red) assuming three years of observation and the detection of one tau neutrino. Also shown are sensitivity prediction for other proposed UHE neutrino experiments (solid lines) and limits on the flux of UHE neutrinos from the Pierre Auger Observatory, IceCube, and Anita (dashed lines). The yellow data points at PeV energies show the IceCube measurement of the astrophysical neutrino flux.}
\label{fig:sensitivity}
\end{SCfigure}

While at least cosmogenic UHE neutrinos are guaranteed to exist, they still have to be observed. Which is why several UHE neutrino experiments with different detection techniques have been proposed in recent years (see \cite{Otte2019a} for references). Here we discuss the air-shower imaging system \emph{Trinity}, which is designed to detect earth-skimming UHE tau-neutrinos. The expected sensitivity of \emph{Trinity} is shown in Figure \ref{fig:sensitivity}. We summarize the characteristics of air showers induced by earth-skimming UHE neutrinos and how they translate into technical requirements for \emph{Trinity} \cite{Otte2019a}. We also present technical solutions that allow to construct and operate \emph{Trinity} by summarizing findings from a white paper that has been submitted to the 2020 Decadal Survey \cite{Otte2019c}.

\section{Characteristics of air showers induced by earth-skimming tau neutrinos}

\emph{Trinity} detects air-showers induced by earth-skimming tau neutrinos. The characteristics of these air-showers differ substantially from that of air showers induced by VHE gamma-rays and UHECRs that start in the upper atmosphere and develop towards the ground. 
An earth-skimming UHE neutrino is one that enters the Earth under an angle of $<10^\circ$. When such a UHE neutrino interacts within the ground a lepton is produced, which is a tau in the case of a tau neutrino. The tau propagates along the projected trajectory of the neutrino and emerges from the ground if it has not decayed before. 

The tau then decays in the atmosphere and starts an upward going air-shower, which develops under the same shallow angle under which the UHE neutrino has entered the Earth. The length over which the shower develops is only 10\,km -- even for a $10^{10}$\,GeV tau neutrino -- because of the higher air density close to the ground. For comparison, a downward going 1\,TeV gamma-ray shower, which is 10 million times less energetic, develops over the same distance in the upper atmosphere and has its shower maximum at an altitude of 10\,km. For the vast majority of tau induced air-showers, the tip of the shower reaches a maximum altitude of 2\,km to 5\,km. 

One way to detect the air shower and reconstruct the characteristics of the UHE neutrino is by imaging the air shower, which is a technique that is very successfully used in the VHE gamma-ray and UHECR communities. To capture an image of an air shower, a sufficient amount of Cherenkov light and/or fluorescence light has to be collected that is emitted while the shower propagates through the atmosphere. 

\begin{SCfigure}[1.0][t]
\includegraphics[width=.6\textwidth]{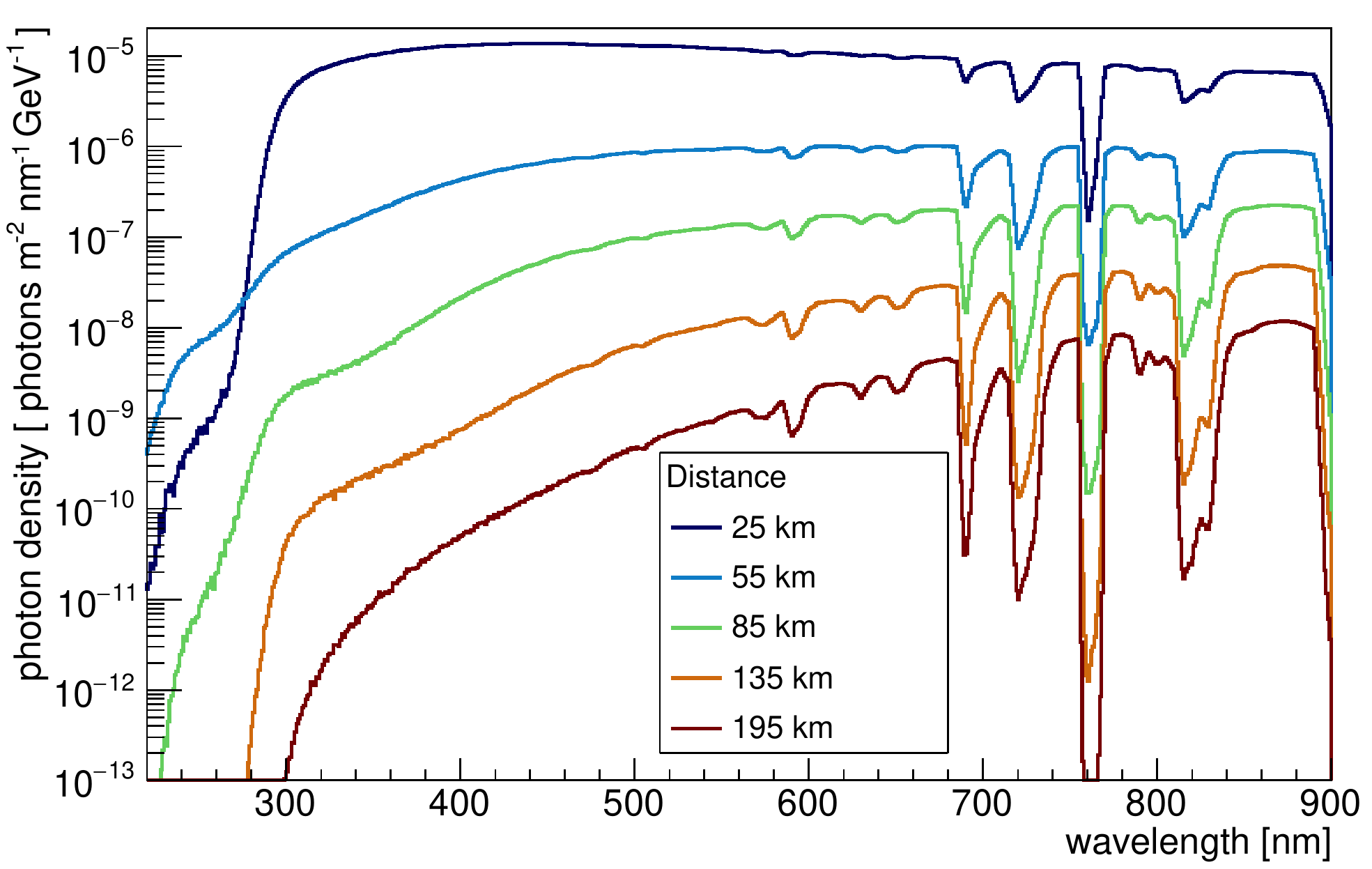}
\caption{Absorbed Cherenkov spectrum simulated for different distances to an air shower induced by an earth-skimming tau neutrino. In the calculation the observer is assumed to be located 1\,km above ground.}
\label{fig:cherspectr}
\end{SCfigure}

\begin{SCfigure}[1.0][t]
\includegraphics[width=.6\textwidth]{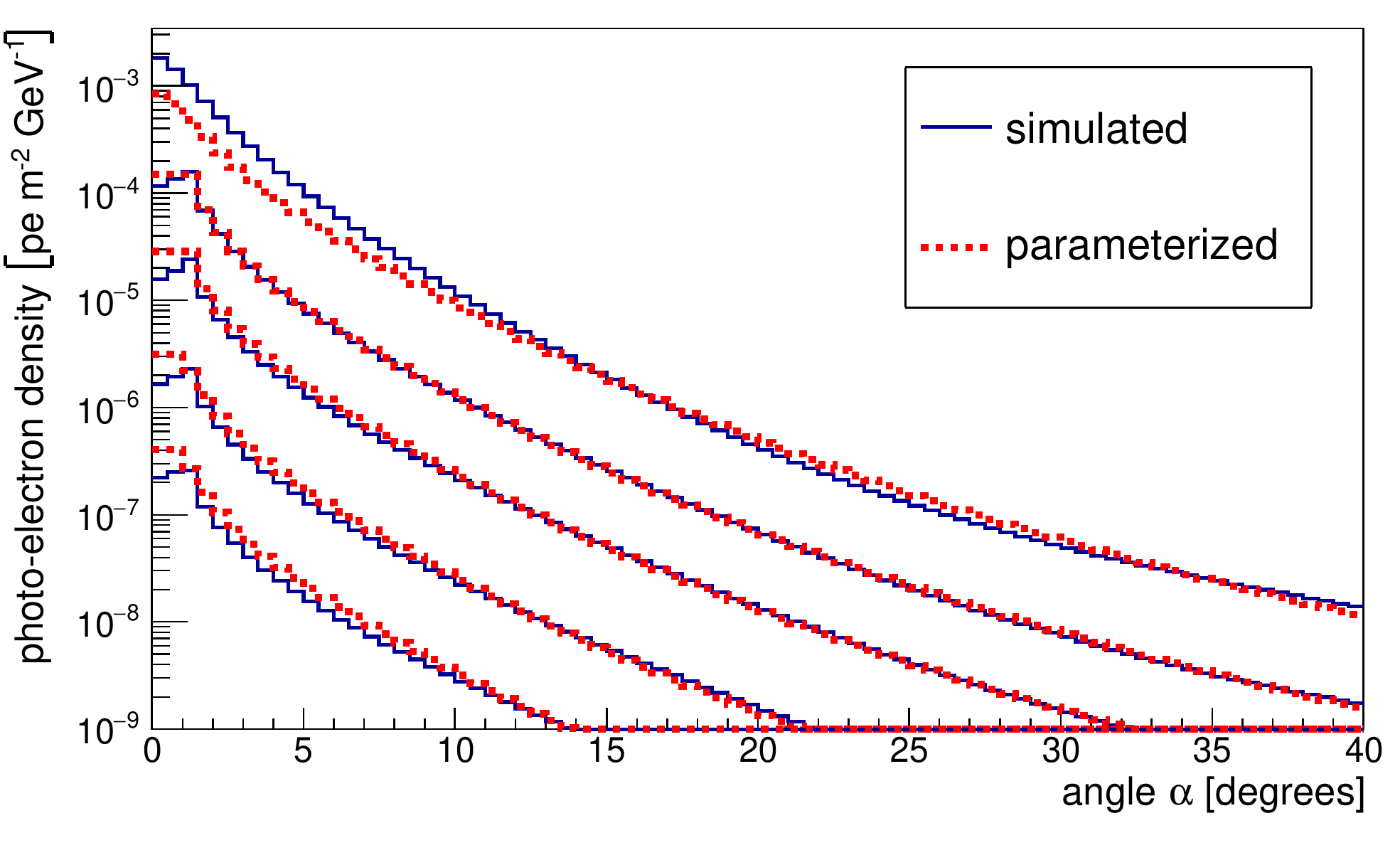}
\caption{Detected Cherenkov intensity per square meter mirror area and GeV shower energy. From top to bottom the distances to the shower are 55\,km, 85\,km, 135\,km, and 195\,km. The intensity is shown as a function of the angle under which the shower is viewed relative to the shower axis. For example, 10 photons will be detected (photoelectrons) from a $10^9$\,GeV shower and per square meter light collection area if the shower is viewed under an angle of $5^\circ$ and the shower develops in a distance of 195\,km, }
\label{fig:cherintensityvsangle}
\end{SCfigure}

Due to a layer of haze, which is typically found in the first one or two kilometers above ground, most of the blue Cherenkov light is absorbed before it reaches the detector. That is illustrated in Figure \ref{fig:cherspectr} where Cherenkov spectra after absorption are shown for different distances to the shower. It is evident that after propagating 100\,km or more, the remaining emission peaks in the red. But despite the strong absorption, the remaining emission is still sufficient to image the air shower. Figure \ref{fig:cherintensityvsangle} shows the amount of Cherenkov light that will be detected with a blue sensitive Hamamatsu SiPM S14520-6050CN per square meter light collection surface and GeV shower energy for different distances to the air shower and viewing angles. 

\begin{SCfigure}[1.0][b]
\includegraphics[width=.6\textwidth]{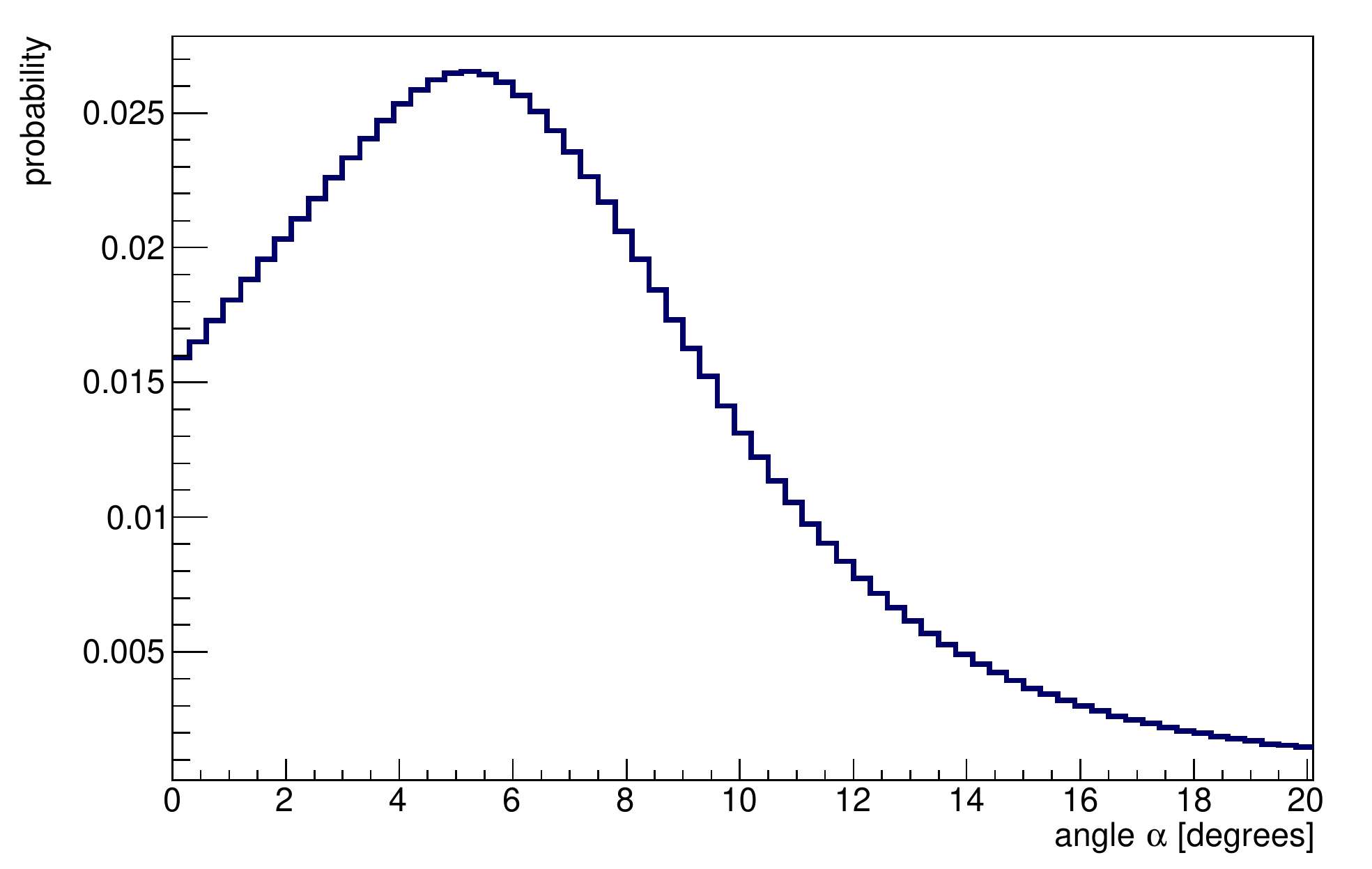}
\caption{Distribution of viewing angles relative to the shower axis of triggered air showers. For this calculation a diffuse neutrino source is assumed with a power-law spectrum and -2 index.} 
\label{fig:triggeredangle}
\end{SCfigure}

From the geometry of the air showers and the intensity of the emitted light, we have derived the baseline configuration of \emph{Trinity} that is discussed in the next section. With that baseline configuration and furthermore adopting a minimum analyzable image length of $0.3^\circ$ and a trigger threshold of 24 detected photons, the distribution of viewing angles of the triggered events peaks at $5^\circ$, with a long tail extending to $20^\circ$ (see Figure \ref{fig:triggeredangle}). See \cite{Otte2019a} for a discussion of the trigger threshold.

\begin{SCfigure}[1.0][t]
\includegraphics[width=.6\textwidth]{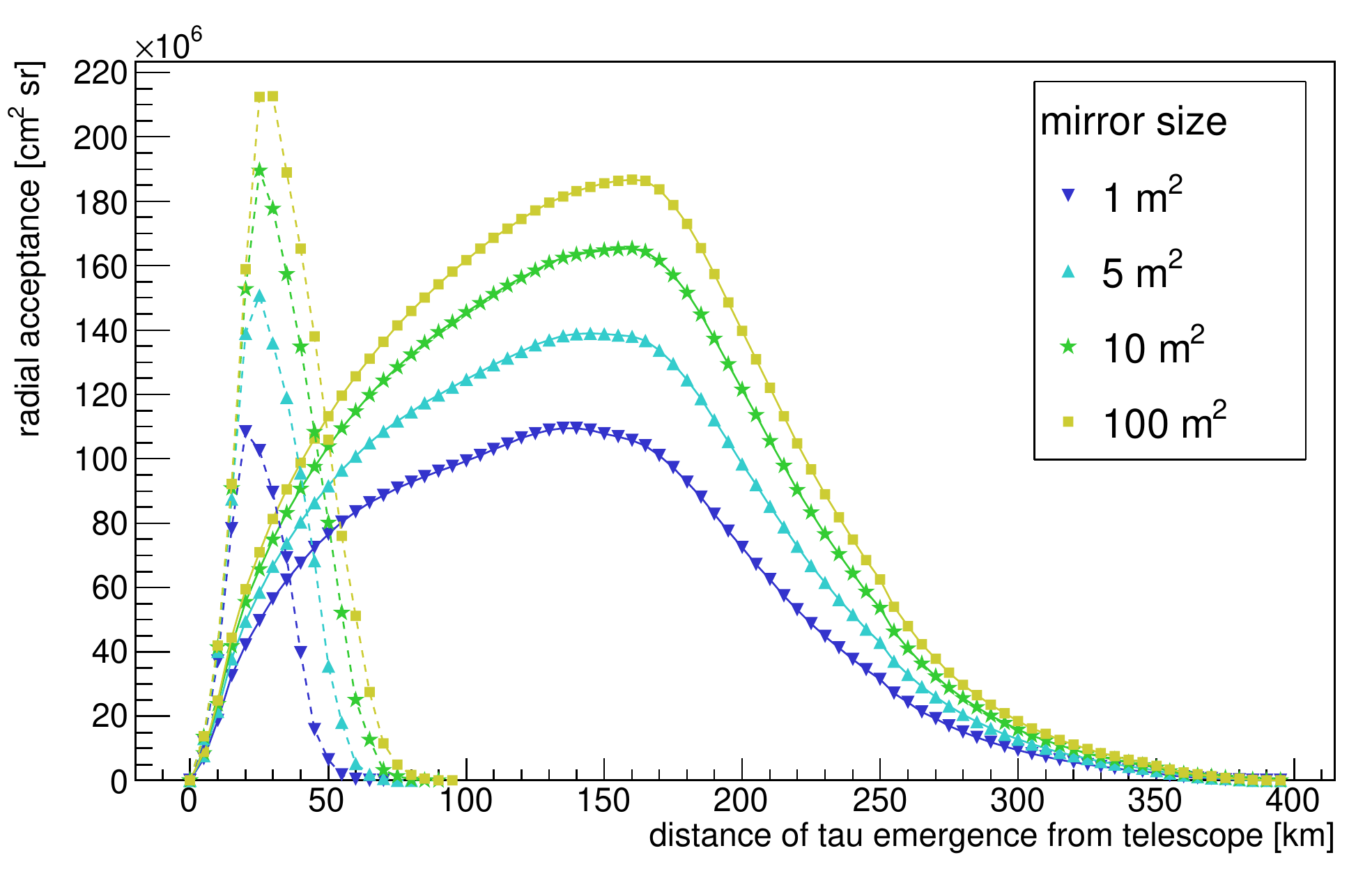}
\caption{Acceptance versus distance from the air shower for four light collection areas. The peak on the left is due to events detected via fluorescence emission and includes about 20\% of all triggered events. The broader peak on the right is due to events detected through Cherenkov emission. The peak of the Cherenkov detected events at $150$\,km coincides with the distance from the telescope to the horizon.}
\label{fig:acceptvsmirrorarea}
\end{SCfigure}

The acceptance of UHE neutrino events with energies between $10^{8.5}$\,GeV and $10^{9.5}$\,GeV for the baseline configuration of \emph{Trinity} is shown in Figure \ref{fig:acceptvsmirrorarea} versus distance from the shower and for four different sizes of the light collection area. It is interesting to note that the acceptance significantly increases up to a light collection surface of $10$\,m$^2$ but only by an additional 13\% when the surface is further increased to $100$\,m$^2$. The acceptance drops for distances larger than 150\,km because a) the Cherenkov intensity drops below the detection threshold and b) an increasing fraction of taus decays and starts developing a shower before the tau is inside the field of view of the instrument. Similar studies have been done for all detector parameters to determine the baseline configuration and are presented in \cite{Otte2019a}.

\begin{SCfigure}[1.0][b]
\includegraphics[width=.6\textwidth]{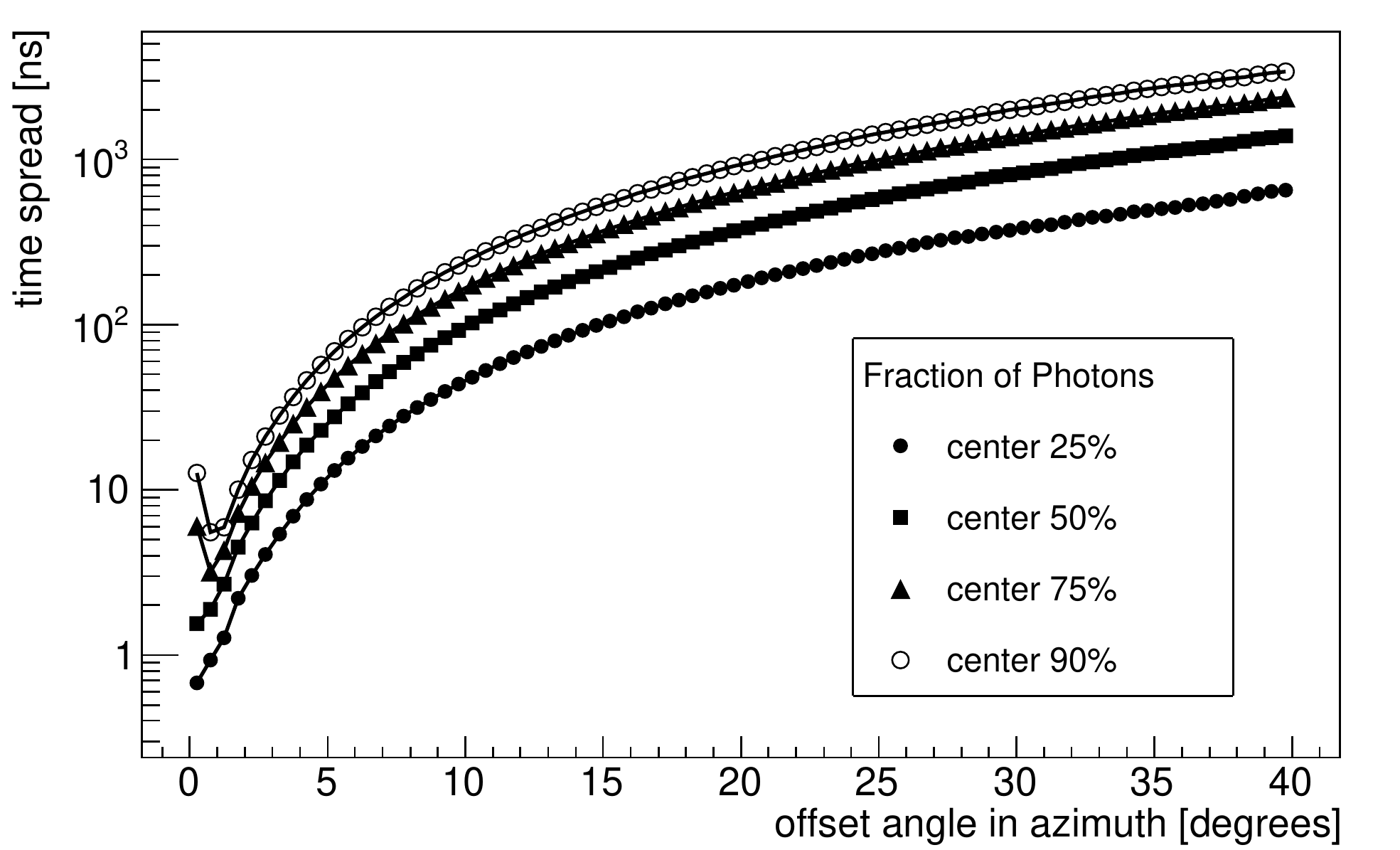}
\caption{Arrival time distribution of Cherenkov photons from a $30$\,TeV electromagnetic shower developing in a distance of 130\,km. The horizontal axis gives the viewing angle relative to the shower axis. The different curves are for different fractions of the Cherenkov photons arriving at the detector.}
\label{fig:cherarrivaltimespread}
\end{SCfigure}

The design of the readout system is influenced by the intensity of the Cherenkov and fluorescence light, the intensity of the background light, and the arrival time distribution of the Cherenkov photons. The latter one is shown in Figure \ref{fig:cherarrivaltimespread} as a function of viewing angle. Because the majority of triggered events are viewed at angles of $10^\circ$ or less, 90\% of the Cherenkov photons of a single event arrive within 10\,ns and 100\,ns. A sampling speed of 100 megasamples per second should be adequate to record the signals with sufficient accuracy. 

\section{Instrument Design}

Based on the characteristics of air-showers induced by earth-skimming tau neutrinos we have derived a baseline configuration for \emph{Trinity} \cite{Otte2019a}. In that configuration, \emph{Trinity} is located at an exposed site 1\,km or higher above the ground. The instrument itself is a system of imaging telescopes, which all point at the horizon with a large field of view that maximizes azimuthal coverage. 

\paragraph{Optics}
The design of the optics for \emph{Trinity} is driven by the requirement to image the entire shower image and do so with with sufficient angular resolution. The containment requirement is fulfilled with a vertical field of view of $5^\circ$ of which $2^\circ$ point below  and $3^\circ$ point above the horizon. Of the $3^\circ$ field of view above the horizon, only $2^\circ$ are needed for image containment. The last degree serves as a veto region to reject downward going air showers from cosmic rays. The angular resolution requirement derives from the desire to analyze images with a length of $0.3^\circ$, which is satisfied with a point spread function of $0.3^\circ$. And finally, the optics has to provide a light collection area of at least 10\,m$^2$ everywhere in the field of view.

\begin{SCfigure}[1.0][t]
\includegraphics[width=.5\textwidth]{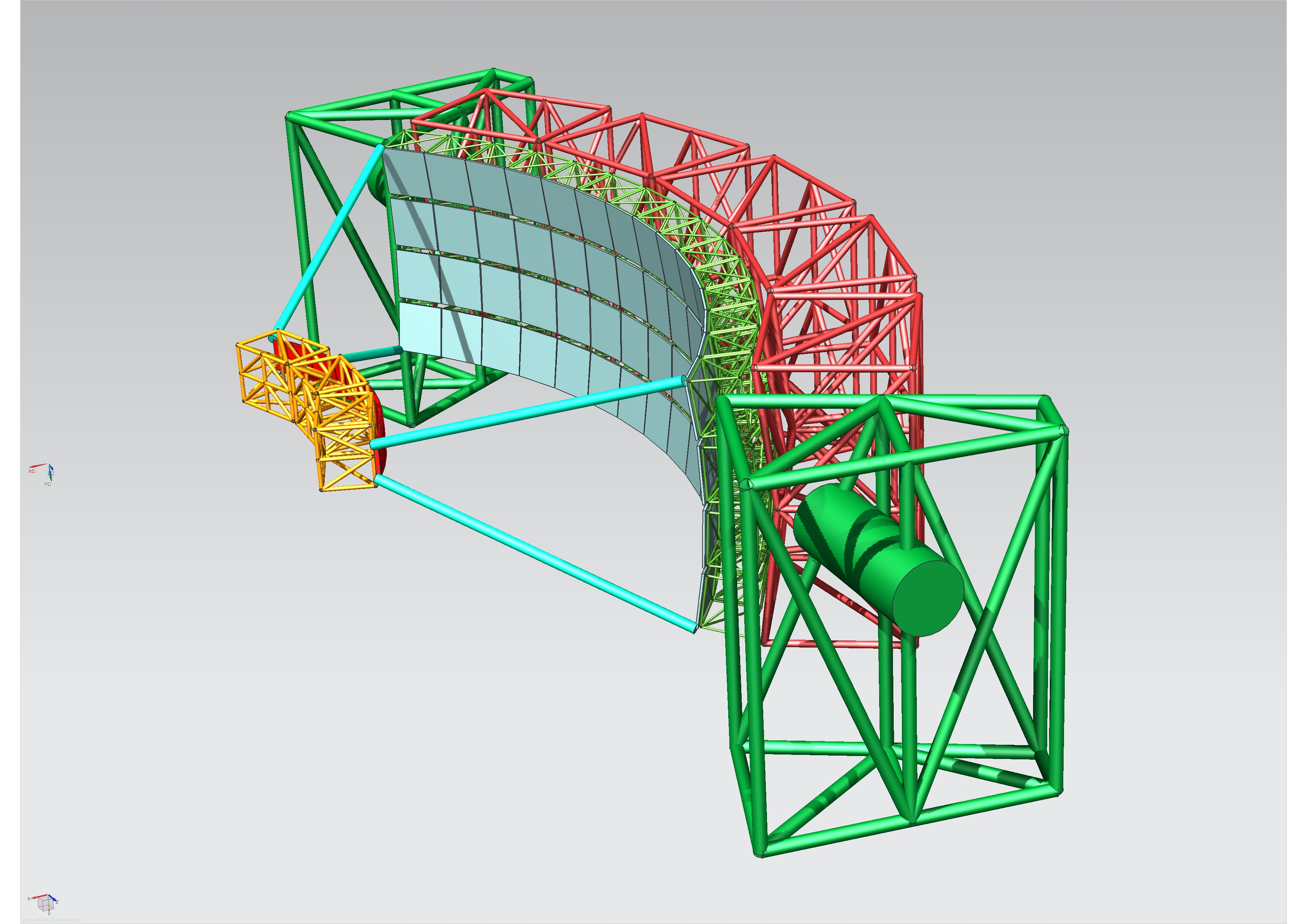}
\caption{Proposed optics for \emph{Trinity} based on the MACHETE optics. The primary mirror is composed of 68, 1\,m$^2$ mirrors. the focal plane (red curved surface) is populated with 3,300 pixels each consisting of a solid non-imaging light concentrator coupled to an SiPM. The field of view covered by one telescope is $5^\circ\times60^\circ$.}
\label{fig:optics}
\end{SCfigure}

All requirements are fulfilled by the optics developed for MACHETE \cite{Cortina2016}. A version of that optics scaled down to provide a $16$\,m$^2$ collection area is shown in Figure \ref{fig:optics}. The system consists of four rows of $1$\,m$^2$ mirrors with 17 mirrors in each row. The technology to built such large mirrors with sufficiently good quality has been developed for CTA. The costs of the mirrors are about \$2k/m$^2$\cite{Otte2019c}.

The focal length of the optics is 5.6\,m and the point spread function is $0.3^\circ$ everywhere in the $5^\circ\times60^\circ$ field of view. Six telescopes will thus provide a $360^\circ$ coverage in azimuth. The focal plane is curved and can be populated with 3,300 pixels, which results in an angular spacing of the pixels of $0.3^\circ$ matching the resolution of the optics. Each pixel is composed of a non-imaging light concentrator made from UV transparent acrylic with an entrance area of 19\,mm$\times$19\,mm coupled to a 9\,mm$\times$9\,mm SiPM. Besides focusing the light onto a smaller sensor area, the purpose of the light concentrator is to reject stray light from outside of the useful mirror area.

\paragraph{Camera and readout}
The photon sensors for \emph{Trinity} have been baselined to be SiPMs, which have a spectral response that matches the red-peaking Cherenkov spectrum. Latest red-sensitive devices from Hamamatsu (S14420-3050WO-Resin) have peak PDEs of $\sim45$\% at 600\,nm \cite{Otte2019b}. For the amplification of the SiPM signals and the adjustment of the SiPM bias voltages, the Application Specific Integrated Circuit (ASIC) MUSIC is a good choice \cite{Gomez2016}. MUSIC was specifically developed to be used in conjunction with SiPMs in air-shower imaging applications. For the digitizer, the AGET system is a viable option, which is a 100 megasamples per second, switch-capacitor-array digitizer system with 12-bit resolution \cite{Pollacco2018}. 

\begin{SCfigure}[1.0][t]
\includegraphics[width=.55\textwidth]{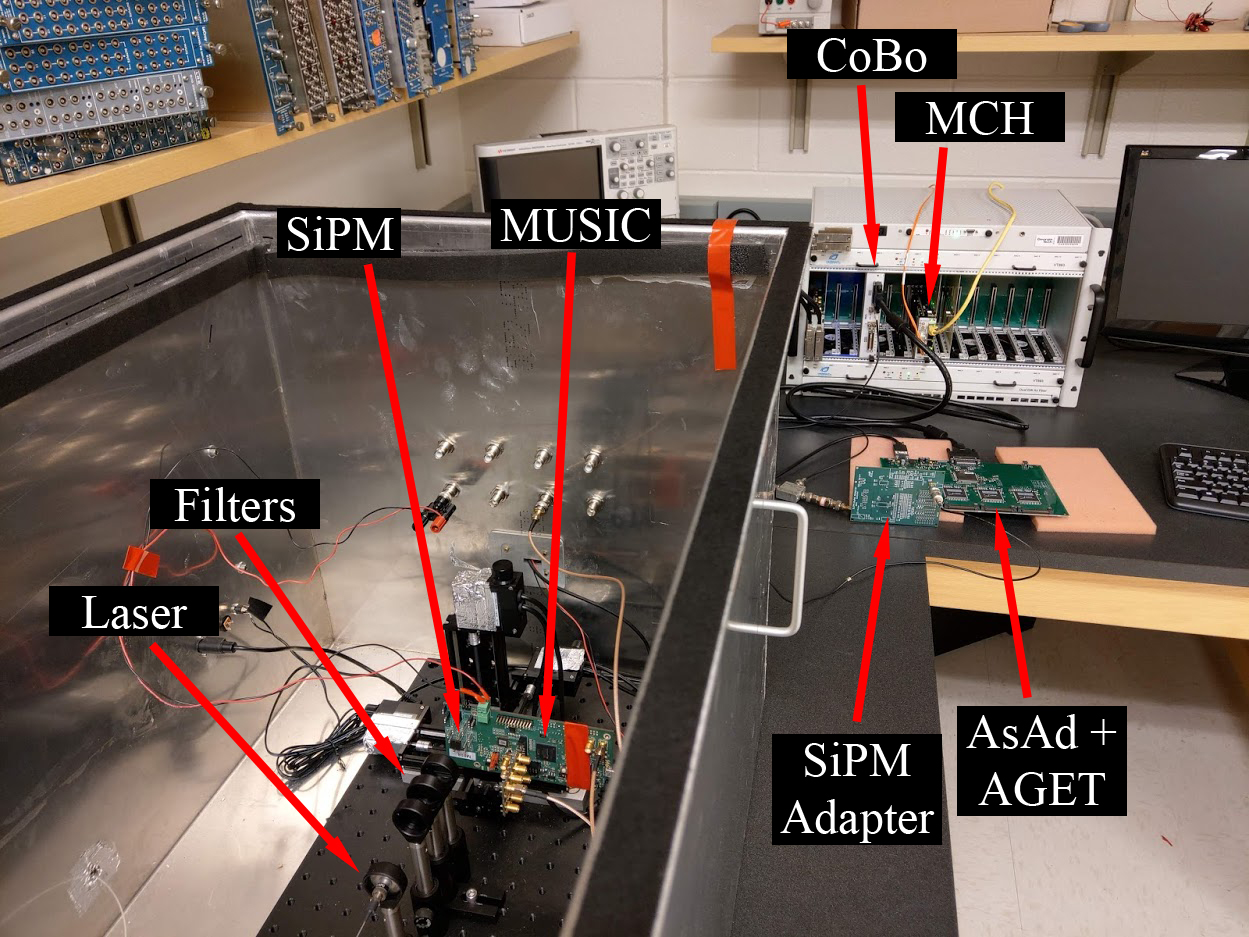}
\caption{Setup to test the EUSO-SPB2 signal chain. A picosecond laser flashes a Hamamatsu S14520-6050CN. The light intensity is varied with calibrated neutral density filters. In the background, the MicroTCA crate with the CoBo and MCH modules are seen, which are needed to communicate with the AGET chips on the AsAd boards and transfer the data to a computer.}
\label{fig:signalchaintestsetup}
\end{SCfigure}

\begin{SCfigure}[1.0][b]
\includegraphics[width=.6\textwidth]{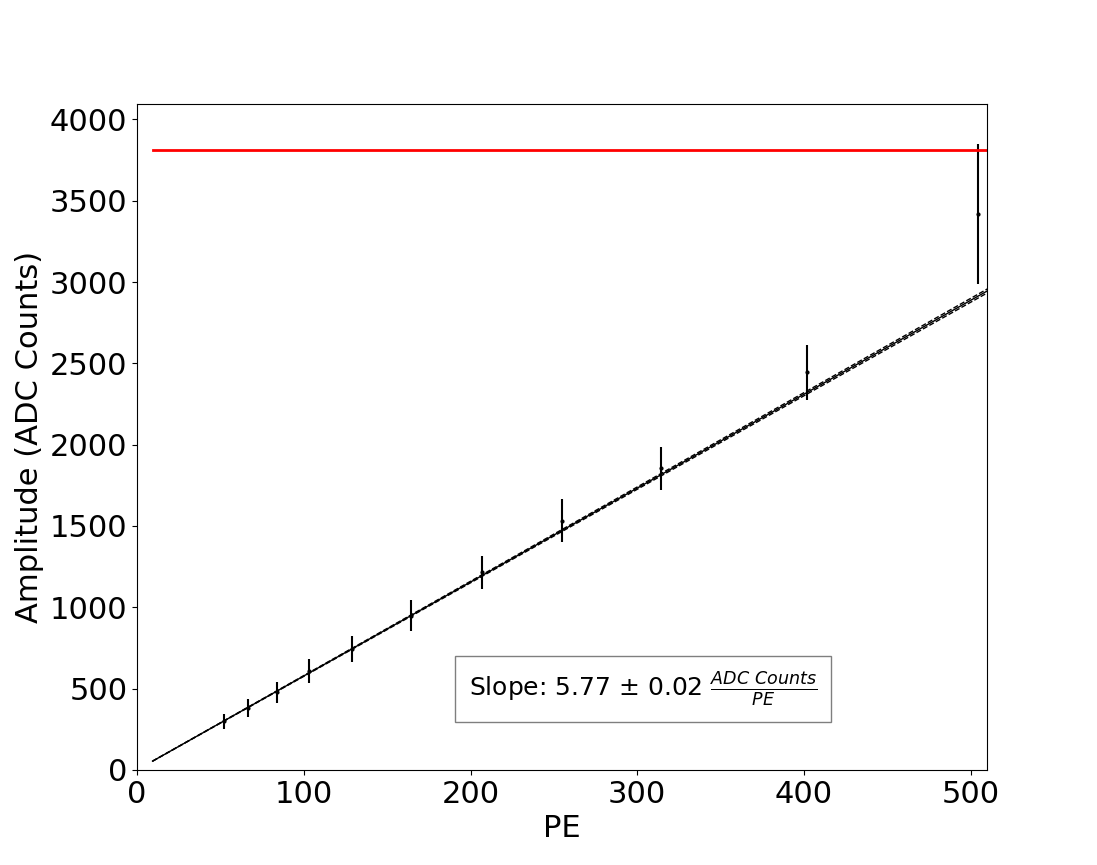}
\caption{Amplitude response of the tested signal chain consisting of a Hamamatsu S14520-6050CN, the MUSIC front-end chip, and the 12-bit resolution AGET digitizer. The response of the system is linear up to an input signal of 450 photoelectrons. The red lines shows the baseline subtracted limit of the digitizer. The MUSIC chip starts to saturate at 450 photoelectrons.}
\label{fig:signalchainamplituderesponse}
\end{SCfigure}

A Cherenkov telescope camera that uses SiPMs, MUSIC, and the AGET system is presently under development for a balloon-borne Cherenkov telescope \cite{Otte2019b}. Figure \ref{fig:signalchaintestsetup} shows the setup to test the signal chain, and Figure \ref{fig:signalchainamplituderesponse} shows the amplitude response of the signal chain. The response of the system is linear up to an input signal of 450\,photoelectrons, which coincides with the maximum dynamic range of MUSIC.

\section{Site}
The geometrical acceptance and thus sensitivity of \emph{Trinity} depends on the accessible azimuthal field of view where \emph{Trinity} is built. The sensitivity in Figure \ref{fig:sensitivity} is achieved if the azimuthal acceptance is $360^\circ$. While it would be advantageous to achieve a $360^\circ$ coverage with one site, the same coverage can also be achieved by deploying telescopes on different sites. In fact, there is no fundamental reason that prevents one from covering more than $360^\circ$ and thus improving the sensitivity by deploying on several sites.

The final site has to feature an ambient background photon intensity that has a minimal contribution from artificial light sources. Less critical are year-round sky conditions; clouds at high altitudes, \emph{i.e.}\ several kilometers above the ground will not affect the duty cycle of \emph{Trinity}. A proper site search has not yet been conducted.

\section{Discussion}

The UHE neutrino band can provide significant insight to contemporary questions in astroparticle physics. The detection of UHE neutrinos, however, is an experimental challenge, which is being tackled by several groups. \emph{Trinity}, a system of air-shower imaging telescopes, is one of these proposed experiments. The sensitivity of \emph{Trinity} reaches well into a region where a viable UHE neutrino flux is expected.

The technology to build \emph{Trinity} has been developed over the past decades mostly by the VHE gamma-ray and UHECR communities. A viable optics concept does exist with the MACHETE design. Mirror technologies that deliver mirrors of sufficient quality have been developed for CTA. \emph{Trinity} greatly benefits from recent improvements of SiPMs, which are now available with peak sensitivities of 50\% in the red overlapping with most of the Cherenkov spectrum. The electronics to operate SiPMs and amplify signals has been greatly simplified with ASICS like MUSIC, which have been specifically designed for air-shower imaging applications. For signal digitization a viable system exist with the AGET. To construct \emph{Trinity}, we estimate a cost of approximately \$5\,M \cite{Otte2019c}.

The construction of \emph{Trinity} can begin with a modest investment in R\&D to finalize the design of the system. Whether the sensitivity shown in Figure \ref{fig:sensitivity} can be reached, depends on the ability to sufficiently suppress background events. While the imaging technique is known to have excellent background suppression capabilities, the observation of events close to the horizon is terra incognita. Addressing this and other potential issues requires the construction and the operation of a prototype telescope station with a partially populated camera. 

\bibliographystyle{pos}
\bibliography{references}

\end{document}